# An iterative approach for generating statistically realistic populations of households

Floriana Gargiulo[1,*], Sônia Ternes[1,2], Sylvie Huet[1], Guillaume Deffuant[1]

1 LISC, Cemagref, Clermont Ferrand, France
2 Embrapa Agricultural Informatics, Campinas - SP, Brazil

*floriana.gargiulo@cemagref.fr

## Abstract

**Background**
Many different simulation frameworks, in different topics, need to treat realistic datasets to initialize and calibrate the system. A precise reproduction of initial states is extremely important to obtain reliable forecast from the model.

**Methodology/Principal Findings**
This paper proposes an algorithm to create an artificial population where individuals are described by their age, and are gathered in households respecting a variety of statistical constraints (distribution of household types, sizes, age of household head, difference of age between partners and among parents and children). Such a population is often the initial state of microsimulation or (agent) individual-based models. To get a realistic distribution of households is often very important, because this distribution has an impact on the demographic evolution. Usual techniques from microsimulation approach cross different sources of aggregated data for generating individuals. In our case the number of combinations of different households (types, sizes, age of participants) makes it computationally difficult to use directly such methods. Hence we developed a specific algorithm to make the problem more easily tractable.

**Conclusions/Significance**
We generate the populations of two pilot municipalities in Auvergne region (France), to illustrate the approach. The generated populations show a good agreement with the available statistical datasets (not used for the generation) and are obtained in a reasonable computational time.

## 1 Introduction

With the increasing computing power, researchers tend to develop models which include more and more diversity and details. A considerable effort has been made, both in academic and corporate research, to generate modelling frameworks simulating policy impacts on complex dynamics: from traffic studies [1] to epidemic diffusion [2,3,4,5], to policy impact studies [6,7,8,9]. These approaches require using various sources of data, detailed at local level to test scenarios with different policies (for instance mitigation strategies) and analyse their impact. For instance, an increasing research effort targets the simulation of epidemic evolution: starting from SARS [3], to the new virus of Influenza A (H1N1) [10]. Many different simulations, at global level or at local level aim at providing precise forecast on the number of infected, with the actuation of different containment strategies. One can expect that such tools become more and more commonly used to support political decisions.

Many models consider populations with an explicit representation of each individual or of the household structures. These individuals are characterised by some state variables (e.g. age, profession, marital status), and often a spatial position. Two main types of modelling approaches can be identified in the literature - Microsimulation and dynamical Individual Based Models (IBMs) [1]: The microsimulation approach defines individual economic and social trajectories through a set of events which occur with given probabilities, generally neglecting interactions between individuals. It provides a mechanism to analyse the effects of policy changes at the level of the decision making units as individuals and households. Individual based models (IBMs) consider the same type of population but generally include more elaborated models of decisions and actions, where individuals take into account the interactions with their environment and other individuals.

In both cases, the dynamics of the whole system is given by the aggregation of all individual behaviours. Hence these modelling approaches are often used to explore the link between the micro and macro dynamics. For instance models of evolving human populations yield demographic patterns in geographical space, which can be compared with census-based data [11].

In both approaches, the first step for the simulation is to initialize the system with a realistic population: the state variables defining the agents or the individuals, must replicate, as closely as possible, the statistical properties of the targeted population. In particular, the demographic evolution must take into account the structure of the distribution of households. Indeed, for the same age structure of the population, different household structures evolve differently.

If individual data were available about the household structure, the problem would be solved quickly by creating a one to one correspondence between the agents and the real persons. However, such a situation rarely occurs, because the institutes managing statistics usually provide aggregated datasets, describing the global properties of the households and individuals. Therefore we must use these aggregate data to generate the artificial sets of individuals and households.

This paper focuses on the specific case of generating a population distributed in households to initialise a dynamical microsimulation model for the PRIMA project. PRIMA – Prototypical Policy Impacts on Multifunctional Activities in Rural Municipalities – is a European project (FP7-ENV-2007) which aims to model the impact of European policies on land use at municipality level in a set of case study regions. Hence in this project, the microsimulation process represents a population of individuals at municipality level, living in households of different types. Once generated, the initial synthetic population evolves through different processes such as birth, death, marriage, divorce, leaving parental house, getting a job and retirement. The quality of the final results depends heavily on the accuracy with which the initial synthetic population represents the available real data.

According to the literature, there are two approaches commonly used to create a synthetic population. In the first approach, some data at individual level are used to create the synthetic population. For instance in the SVERIGE model [9], the whole population of Sweden in 1990 is the starting population, and large longitudinal data sets are used for estimation of many equations for the demographic process. In a similar way, DYNAMOD [12] is a dynamic model designed to project population characteristics over a 50-year period, using a 1% sample. A second approach uses the Iterative Proportional Fitting [13] to estimate the joint probability of characteristics belonging to different sets of aggregated data. This approach is used in the SMILE model [6] where the synthetic population is generated from Census of Small Area Population Statistics (SAPS) in 1996 in Ireland, considering characteristics as gender, age, employment status and industry, for a given group of the population in a specific location. IPF can be applied when the Census data, describing the aggregate properties of individuals and households, are integrated with individual

---

[1] Sometimes they are also called "agent based models", because the individuals represent economic or social agents. But there is an ambiguity with a different research trend of "agent based models", more related to computer science, which investigates computer agents that cooperate for achieving some tasks, for instance foraging on the internet. To avoid this ambiguity, we prefer to use the expression "Individual Based Models", which originally comes from modelling in ecology.

data, extracted by surveys on samples that can be bigger or smaller than the size of the desired artificial population. Thus, the initialization process consists in finding the good weight to attribute to each sub-element of the analyzed sample to make it representative of the objective population. Some methods to solve the up-scaling or downscaling initialization problem, with stochastic and deterministic approaches, are described in [14,15,16,17,18]

In our problem, individual data to cross with the aggregate properties are not available. This situation does not allow us to apply the IPF method. Moreover computing the joint probability of characteristics of households, including size, type and age of members, implies heavy computations. In this paper, we propose an iterative semi-stochastic algorithm, involving a sequence of stochastic extractions, which considerably decreases the computational cost of the population generation. This algorithm uses only aggregated datasets from Census, and the missing crossings between the data are obtained through testing procedures.

The algorithm is adjusted for data from the Auvergne region (France), but the general concept can be easily adapted to different uses. The next section describes the details of the problem to solve. Section 3 describes the available data in Auvergne region, as well the attributes of the synthetic population to be generated. The iterative algorithm is described in detail in section 4. Sections 5 and 6 present the results and conclusions.

## 2 Materials and Methods

### 2.1 General Formulation of the problem

The classical generation approach only considers one micro level (individuals or households). The specificity of this work is that we need to respect statistical constraints on the distribution of the individual ages, the distribution of household size and the distribution of individual ages within households.

More precisely our problem is to generate a set of households comprising individuals taken in a distribution of age of the population, and which respect all the constraints we found in the data about the distributions of:
- sizes and types of households,
- ages of the head of the household,
- differences of age between partners,
- ages of children according to mother's age.

Let us call:
- $t$ the type of household, the values of $t$ can be: 'single', 'couple', 'single-parent', 'complex';
- $s$ the size of the household, the values of $s$ can be: 1, 2, 3, 4, 5, >5;
- $a_r$ the age of the head of the household;
- $a_1, \ldots, a_{s-1}$ the age of the children of single-parent households;
- $a_{r'}$, the age of the head's partner, and $a_1, \ldots, a_{s-2}$, the age of the children for couple households
- $(a_i)$ generally represents the list of the ages of the household members.

In a first approach we would suppose that we are able to compute a good approximation of the probability of a given household $P(t, s, (a_i))$ (a possible method to compute these probabilities is described in section 2.3). Then, a straightforward way to proceed is described in algorithm 1.

**Algorithm 1:**

1. Generate all possible households, considering all possible combinations of types, sizes and ages of members;
2. Associate with each of these households, defined by the values of (*t*, *s*, ($a_i$)), the probability $P(t, s, (a_i))$;
3. Generate a void list *H*. Repeat, until the size of *H* reaches the expected number of households:
   a. Pick a household generated in step 1 according to its probability associated in step 2;
   b. Add the household to list *H*.
4. Return *H*.

This algorithm shows a significant drawback. Although the average on a large runs of this algorithms of the distribution of age will be close to the data, one can expect significant differences between the age distribution of a specific run and the data. Since the data about the distribution of ages are reliable in our problem, we would like to keep it as precise as possible in our approach.

This leads to algorithm 2, where we use the list of ages of individuals directly taken from the data, and a probability of household $P'(t, s, (a_i))$, independently from the distribution of ages in the population:

**Algorithm 2:**
1. Generate a population of individuals following the age structure of the population. Let us call it the list $I = \{ a_j \}$ (to each element of the list an age is associated);
2. Generate all possible households, considering all possible combinations of types, sizes and ages of members;
3. Associate with each of these households, defined by the values of (*t*, *s*, ($a_i$)), the probability $P'(t, s, (a_i))$ of the household, independently from the age distribution of the population;
4. Generate a void list *H*. Repeat, until list *I* is void or a number *N* of iterations is reached:
   a. Pick a household *h* generated in step 2 according to its probability $P'(t, s, (a_i))$;
   b. If ages ($a_i$) are included in *I* then remove them from *I* and copy household *h* in *H*.
5. Return *H*.

With algorithm 2, we guarantee to keep the final distribution of individual ages close to the data. Generating the list of individuals following the age structure of the population is straightforward. The Census data of 1990 [19], the starting point at which we initialize the model for the Auvergne region, chosen in the PRIMA project as a pilot region to be studied, provides the age distribution for the population at municipality level. Two municipalities are chosen to test the algorithm: Abrest, which was composed by 964 households with a total population of 2545 individuals, and Bellerive-sur-Allier, composed by 8530 individuals organized in 3520 households. The choice of these municipalities was made arbitrary, considering the difference of sizes, for testing the algorithm.

These data, displayed in Figure 1, allow us to generate directly the list *I* of individuals following the age structure of the population. Simply, going through all the age brackets, and for each one, we add to the list the corresponding number of individuals.

However, the other steps of the algorithms involve several difficulties:
- To evaluate the probability of a given household. This will be addressed in section 2.2.
- To manage the complexity of the set of all possible households. This will be addressed in section 2.3.
- In general, the algorithm leaves some individual ages unused at the end, and generates a smaller number of households than expected (because of the impossibility to find the necessary individuals to fit the drawn households). This is also addressed in section 2.3

## 2.2 Calculating the probability of a household

Census data, [19], provide also some information about households: the size distribution, the age distribution for people living alone (single households) and the age distribution of the head of the household. Figures 2 to 4 show those available data for the two municipalities. From those data, we can calculate the probability of each household.

Data of figure 2 provide us with $P(s)$, the probability of having a household of size $s$.

Data of figure 3 provide us with $P(a_r | s=1)$, the probability of age range of the head for households of size 1 (single).

Data of figure 4 together with data of figure 3, provide with $P(a_r | s>1)$, the probability of age range of the head for households of size superior to 1.

Data of figure 5 provide us with $P(t | a_r = \alpha)$, the probability of a household type given the age of the head equals $\alpha$, and the probability $P(\text{child} | a = \alpha)$ for a individual of age $\alpha$ to live in a household without being the head or the partner (this means, as a "child"[2]). Involving this constraint is very important to avoid to get households with very old parents (e.g. 90 years) and old children (around 70).

Clearly these data at local level are not sufficient to characterize a household. We lack constraints on the distribution of ages inside a given type of household. Hence we used some data at national level about the age structure inside the households regarding the ages of parents on one hand, and the ages of children on the other hand. Figures 6 and 7 show the national level data that we use to calculate the probability of the structure of ages, [20,21].

Data of figure 6 provide us with $P(a_{r'} | a_r = \alpha)$, the probability of the age of the head's partner, given the age of the head.

From data of figure 7, we can derive $P((a_i) | a_m = \alpha, s = \sigma)$ the probabilities of children ages knowing the number of children and that the age of the mother is $\alpha$. We consider that in couple households, the mother is the partner, and in single-parent households, the head is the mother.

We can now use these partial probabilities to evaluate the probability of a given household $P(t, s, (a_i))$. We must distinguish cases 'single', 'single-parent', 'couple':

$P'(t=\text{'single'}, s, a_r) \quad = P(s=1) * P(a_r | s=1)$
$P'(t=\text{'singlep'}, s, a_r, (a_i)) \quad = P(s=\sigma) * P(a_r | s>1) * P(\text{'singlep'} | a_r) * \Pi\, P(a_i | a_r)\, P(\text{child} | a_i)$
$P'(t=\text{'couple'}, s, a_r, a_{r'}, (a_i)) \quad =$
$\qquad P(s=\sigma)) * P(a_r | s>1) * P(\text{'couple'} | a_r) * P(a_{r'} | a_r) * \Pi\, P(a_i | a_{r'})\, P(\text{child} | a_i)$

This evaluation theoretically allows us to apply the approach of algorithm 2. However, to generate all the combinations of households and picking one according to these probabilities is computationally expensive. In the next section, we propose an iterative algorithm which is more efficient computationally.

---
[2] That is the definition of "child" for the French Census managed by INSEE

## 2.3 An iterative algorithm avoiding to generate all possible households

The principle of the algorithm is to build progressively the household, by picking its member(s) according to the previously described probabilities, and, for each new member, to test if there is an individual of this age in the list of individuals *I*. If not, we stop the process for this household and begin to build another one.

The flux diagram describing the process is represented in Figure 8.

The algorithm consists of five main steps (see algorithm 3).

**Algorithm 3**
1. Pick the size of the household according to $P(s)$;
2. Pick the age range of the head according to $P(a_r|s)$. If there is no individual in *I* of the age range, the process is stopped and a new attempt for building a household is launched. Otherwise an individual of the chosen age range is added to the household, and removed from list *I*;
3. If $s > 1$, pick a household type ('couple' or 'single-parent') according to $P(t|a_r)$. "Complex" households are not considered at this stage.
4. If $t$ = 'couple', pick the age of the partner according to $P(a_r'|a_r)$. Again, if there is no individual in *I* of the chosen age range, then the household is abandoned, the head is put back to list *I*, and a new attempt to build a household is launched. Otherwise an individual of the chosen age range is added to the household and remove from list *I*;
5. We pick the age of children with probability $P(a_i | a_r)*P(\text{child} | a_i)$ for single-parent and $P(a_i | a_r')*P(\text{child} | a_i)$ for couples. Again, for each child, if there is no individual in *I* of the chosen age range, then the household is abandoned, its members put back to list *I*, and a new attempt to build a household is launched. Otherwise an individual of the chosen age range is added to the household and removed from list *I*.

This process is equivalent to pick one household according to its evaluated probability, and keeping it if all the ages of its members are present in list *I*. Indeed, the process of picking the different members of the household leads to the same overall probability to pick a household, and since the attempt is cancelled as soon as one age is lacking in list *I*, it changes nothing to make these tests iteratively.

Moreover, we can constrain even more the process by considering the list of household sizes which is directly derived from the data. The rest of the process remains the same. Then we are sure to have the right number of households, even though when algorithm 3 stops, some void households remain in the list.

Indeed, the described algorithm should *a priori* be repeated until all the households of the list are filled with all the individuals of the availability vector. However, this situation is never reached and after the creation of almost all the households, the program reaches a point where no more households can be achieved given the remaining individuals. For this reason, when this situation is reached, the algorithm is stopped. The remaining households can be considered as "complex structures", namely all the housing solutions that cannot be placed into the usual categorization of household type (single, couple, single-parent). A complex household can be, for example, a group of students occupying the same dwelling or two familiar groups sharing the same location. Therefore, since we do not have any information about these structures from the data sets, to conclude the generation of the artificial population, the complex households are filled randomly with the remaining individuals in the availability list.

# 3  Results

We tested the algorithm for two different municipalities in Auvergne: Abrest and Bellerive-sur-Allier. The first one had a population of 2545 inhabitants in 1990, while the second one had 8530. In the following we compare the statistical properties of the artificial population with the real Census data. We use for the comparison both the data implicitly used in the building algorithm and other national and municipality level data, which were not used in the generation process. We calculate the distributions both for one single realization of the system and for a sequence of 100 realizations (the random nature of the algorithm leads to some variations from one run to the other).

By construction of the algorithm, the age distribution and the size distribution of the household are directly derived from the data for the two villages. In Figure 9 we show the distributions of the age of head for real data and the artificial population. The distribution of the age of head was used inside the generation process, but the stochastic extractions from this distribution were spaced out from various tests; for this reason we can expect some discrepancy between the real data and the generated population.

As we can observe in Figure 9, the artificial population respects quite well the real distribution.

In Figure 10, we compare the obtained artificial population with the real distribution of number of children in households. This particular data set was not used in the generation, so the comparison can give an idea of the accuracy of the algorithm; this data set is reported in Table 1. Also in this case we can observe a good agreement between the real data and the simulations.

The final comparison (Figure 11) regards the household typology. For this comparison we will not use directly the data that we have used in the generation (the probability for a person to be in a certain type of household) but another dataset containing the direct proportions of household types at national level. This dataset is reported in Table 2.

In this case the differences from the real data, for both municipalities, are quite significant. It could be expected: the data we are using in this case for the comparison are at national data, and therefore keep into account of the population of metropolitan areas. The discrepancy between our results and the national data, therefore, do not highlight an error in the generating process, but show the behavioural difference between metropolitan area and rural villages, with small population.

Moreover, it is noticeable that the data reported in the previous graph provide relevant information about the complex households. We lack completely this information at village level and therefore we cannot use any constraint on complex households in the building procedure. In the proposed algorithm, complex households are created randomly, grouping together the individuals that the generating procedure cannot assign to a household according to the selection/test mechanism. Nevertheless, we observe that the proportion of complex households that we obtain is close to the data at national level.

Finally, we need to stress that this kind of algorithm is strictly correlated to the data structure we have: for Auvergne region such as for France and most of occidental countries the main household structures are based on the concept of "nuclear family": a couple of parents and a certain number of children, or a subset of this structure. In some other cultures the basic household can have completely different structure (for example many generations sharing the same dwelling), and therefore this kind of approach can give rise to potential bias without any additional information about the structure of complex households.

# 4   Conclusion

In this paper we proposed an algorithm for the generation of a synthetic population organized in households that can be applied in various modeling contexts. This method gives good results without using a set of prototypical households that, in many cases, are not available. This is an advantage compared with existing methods such as IFT. Moreover it allows one to reproduce exactly some features of the real population that are particularly important for the subsequent analysis.

This algorithm is a practical implementation of a general approach where the households are picked according to their probability, among all the possible household structures. The method builds the households iteratively. It tests the availability of the age of its members at each step, and backtracks as soon as an age is lacking. This saves a lot of computations.

We presented the example of the PRIMA project, where the artificial population is needed as initialization of a dynamical microsimulation model at municipality level. We showed that the algorithm yields a good agreement between the statistics of the artificial population and the real one. Clearly, the approach can be adapted to other cases where it is necessary to generate a population organized in households. During the project, we shall have to adapt it to other sets of data that can be found in different case study regions.

The algorithm can deal with other properties of the individuals and of the households. For instance, we could add a gender variable to describe the individuals of our example. We would need to split the list *I* of individuals of different ages into two lists, one for males and one for females. Moreover, we would need to include the percentage of household where the head is a male and about the percentage of heterosexual couples. Then the principle of the method remains the same. The only difference is that to build the households, we pick either in the list of males or in the list of females.

More generally, after the set-up of the demographical structure, other characteristics can be assigned to each individual, through stochastic extractions or deterministic associations: the level of instruction, the professional activity, the favorite recreational activities, the commuting pattern, etc. According to the available datasets, these properties can be assigned to each individual independently from the household in which it is embedded, or some correlations can be considered inside the same household.

**Acknowledgment**
This publication has been funded under the PRIMA (Prototypical policy impacts on multifunctional activities in rural municipalities) collaborative project, EU 7th Framework Programme (ENV 2007-1), contract no. 212345.

**Figures**

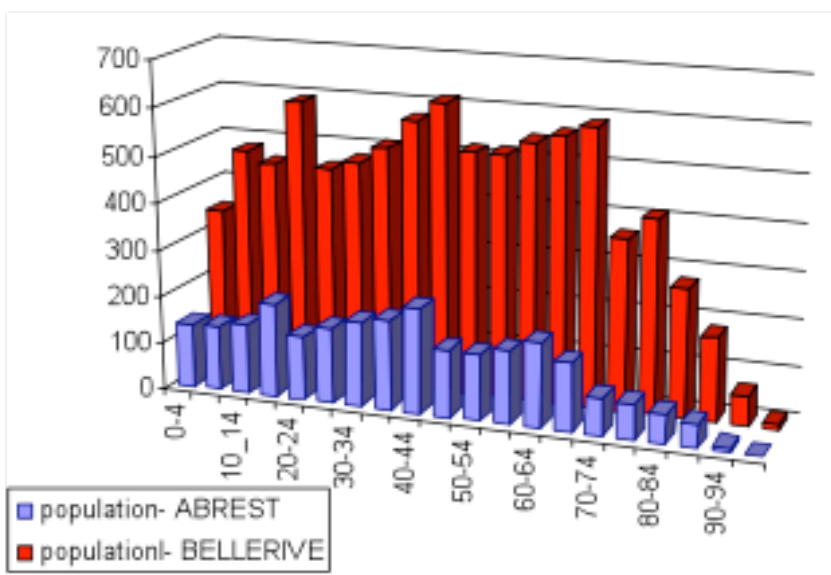

**Figure 1**: Histogram of the number of individuals according to various age ranges of 5 years each for Abrest and Bellerive-sur-Allier. Source: INSEE, French Census data, 1990.

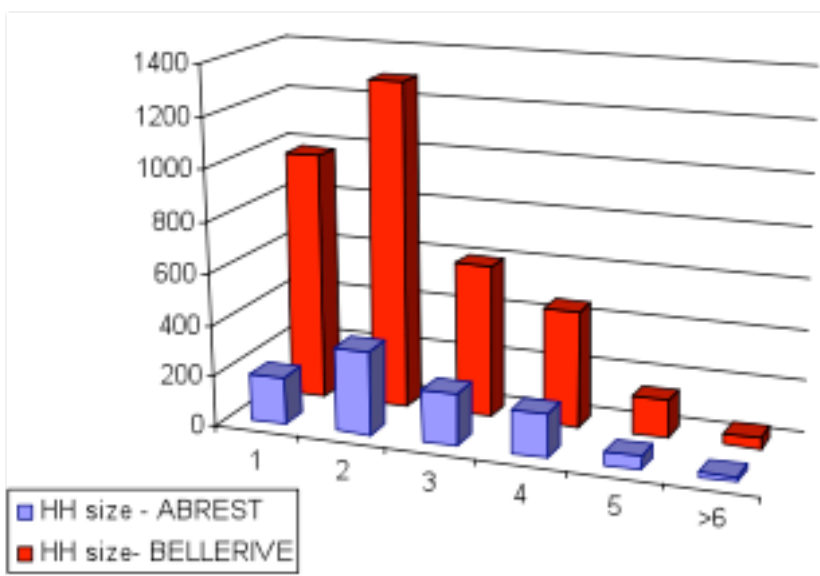

**Figure 2**: Histogram of the number of households according to their size (number of individuals in the household) in Abrest and Bellerive-sur-Allier. Source: INSEE, French Census data, 1990..

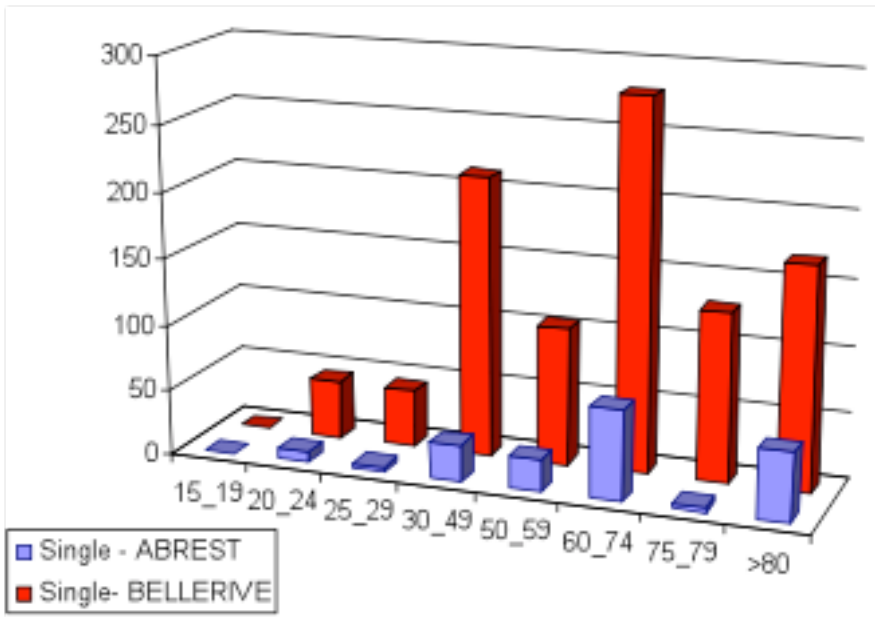

**Figure 3**: Histogram of the number of households according to the age ranges of person living alone in Abrest and Bellerive-sur-Allier. Source: INSEE, French Census data, 1990..

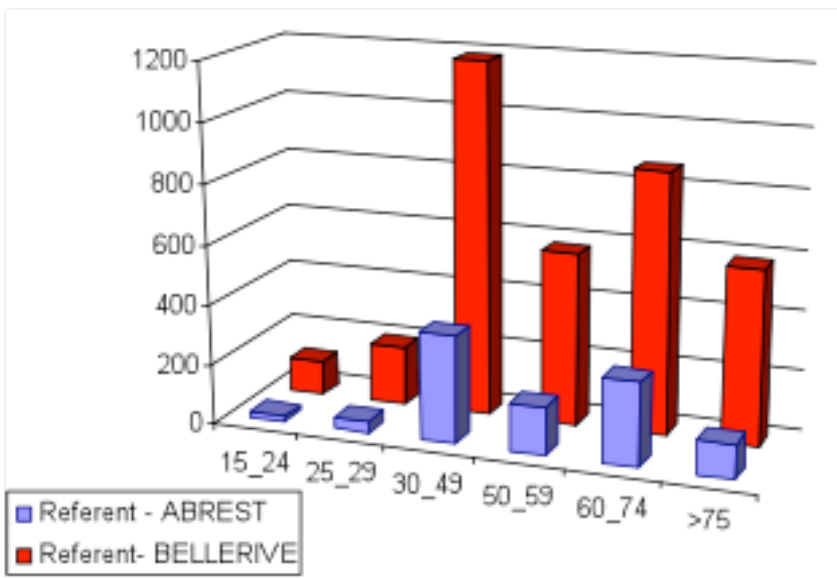

**Figure 4**: Histogram of the number of households according to the age ranges of the head in Abrest and Bellerive-sur-Allier. Source: INSEE, French Census data, 1990.

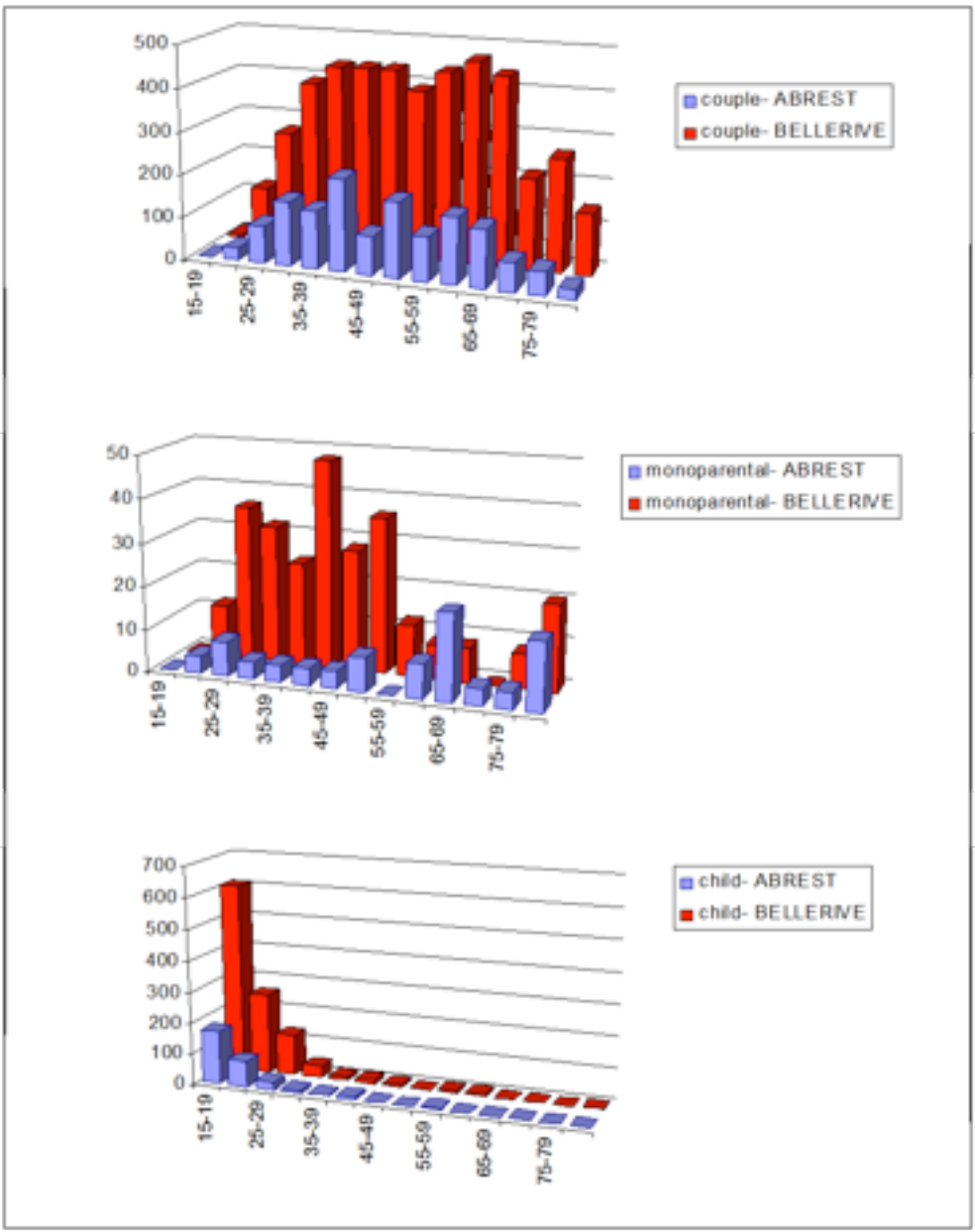

**Figure 5**: Histogram of the number of individuals of age > 15 according to different age ranges, from the top to the bottom, living as partners in couple, as head in single-parent households or living with parent(s) in Abrest and Bellerive-sur-Allier. Source: INSEE, French Census data, 1990

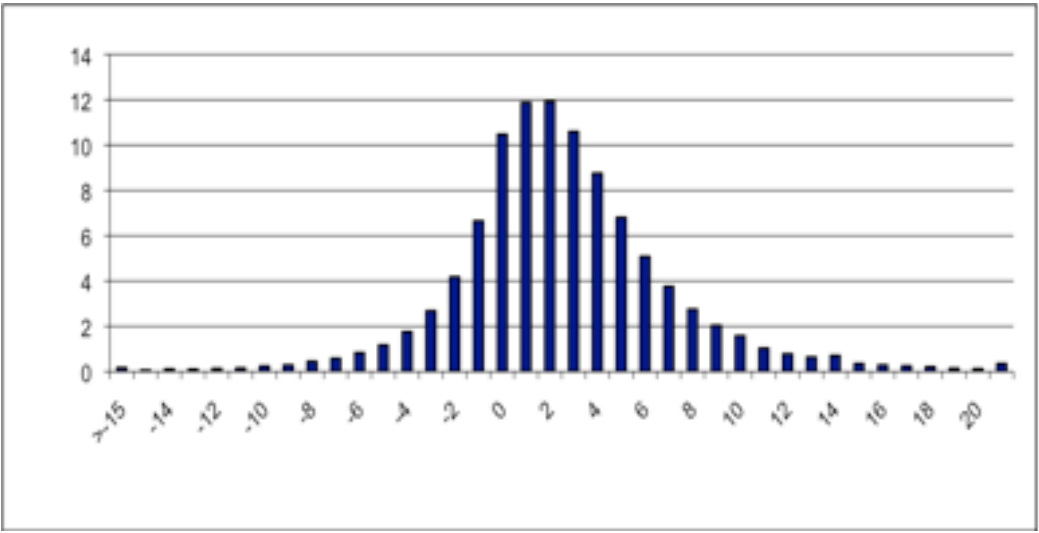

**Figure 6**: Histogram of the number of couples according to their difference of ages in France in 1999. Source: INSEE, "Enquête sur l'étude de l'histoire familiale de 1999".

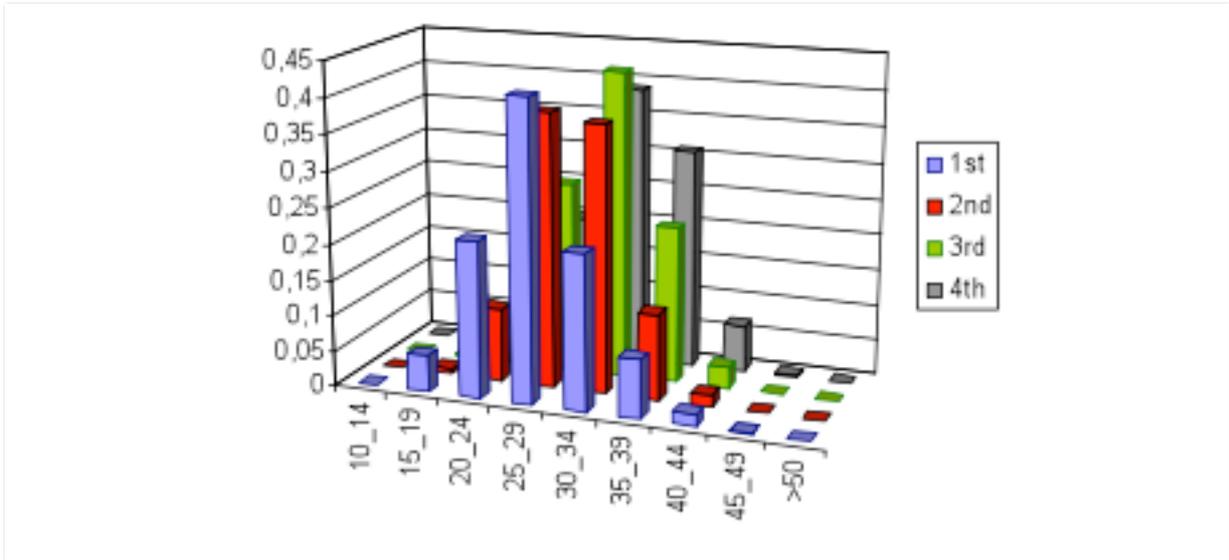

**Figure 7**: distribution of live births by birth order and mother's age range in France. Source: Eurostat Data 1999.

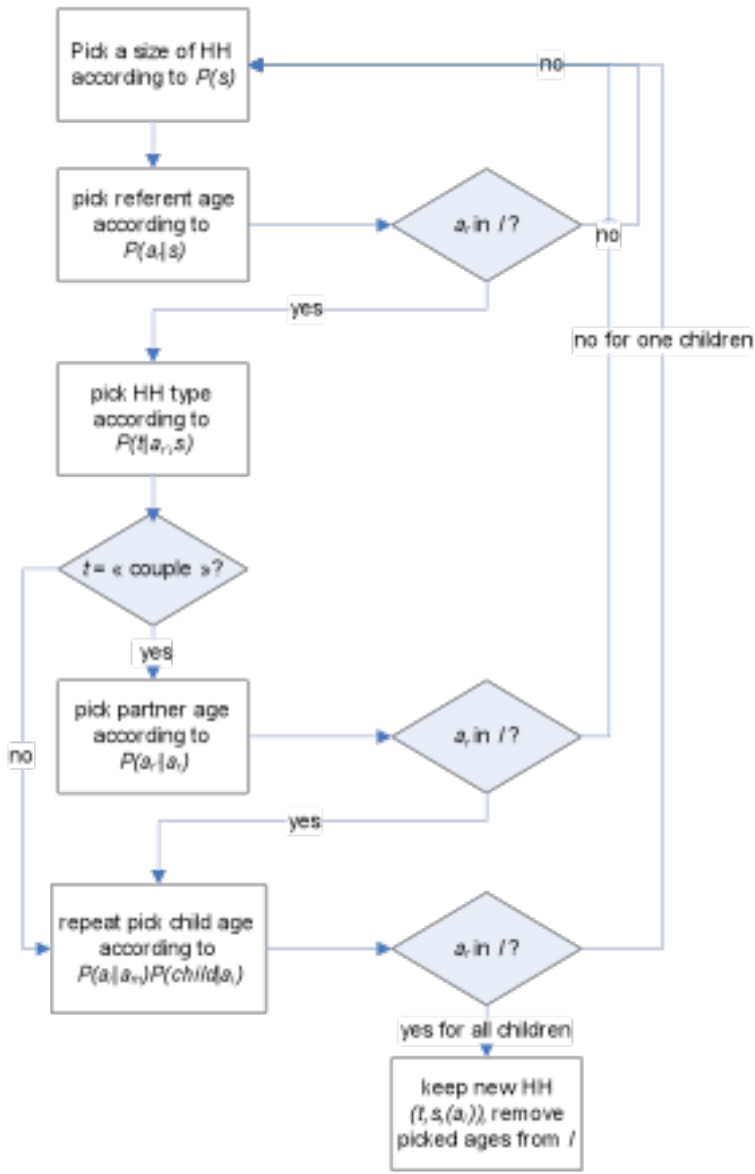

**Figure 8**: Flux diagram describing the algorithm for the generation of an artificial population for PRIMA project

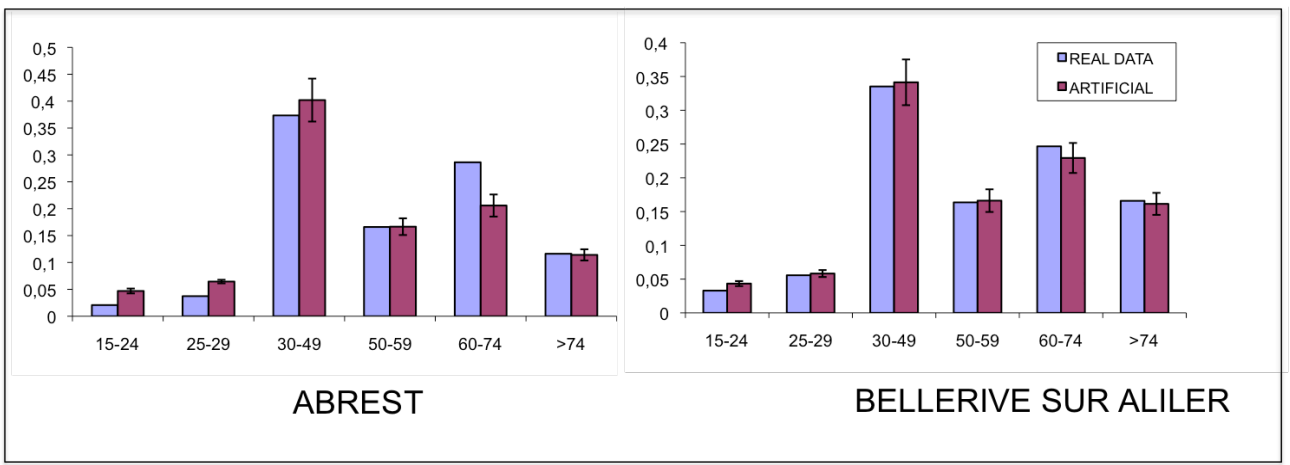

**Figure 9:** Histograms for age of head distribution for the municipality of Abrest (left plot) and of Bellerive-sur-Allier (right plot). The light purple bars represents the real data, the dark purple bars the average for 100 realizations for the artificial population. The error is the standard deviation on the 100 replicas.

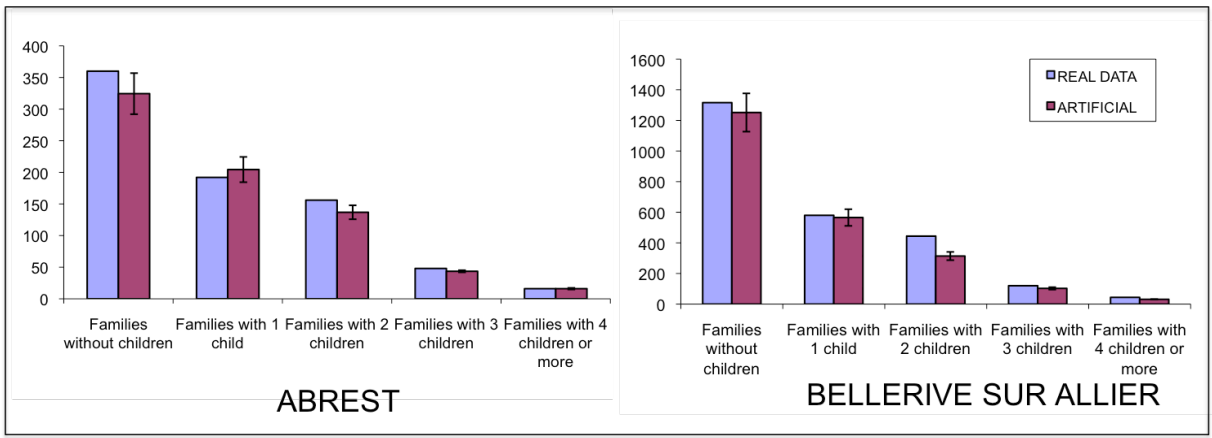

**Figure 10:** Histograms for age number of children distribution for the village of Abrest (left plot) and of Bellerive-sur-Allier (right plot). The light purple bars represents the real data, the dark purple bars the average for 100 realizations for the artificial population. The error is the standard deviation on the 100 replica.

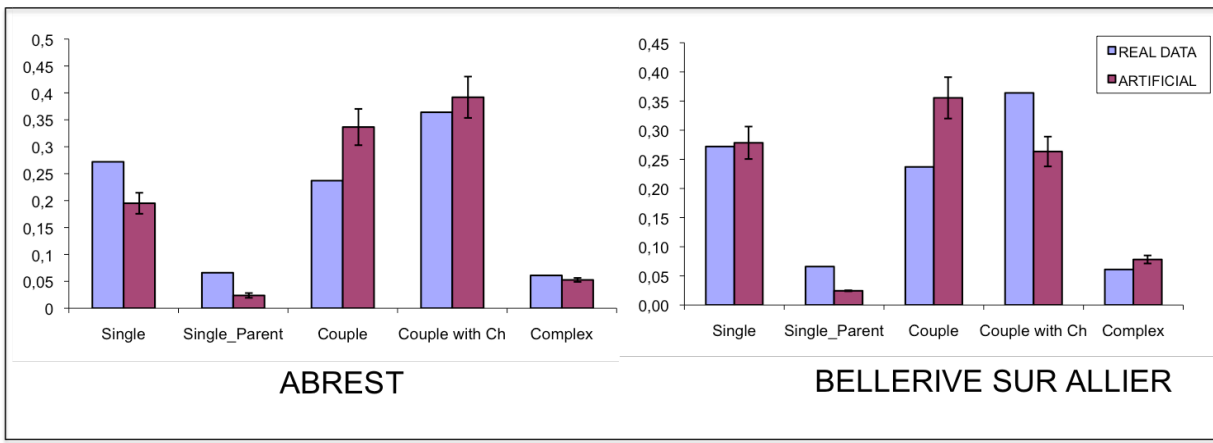

**Figure 11:** Histograms for the household type distribution for the village of Abrest (left plot) and of Bellerive-sur-Allier (right plot). The light purple bars represents the real data, the dark purple bars the average for 100 realizations for the artificial population. The error is the standard deviation on the 100 replica.

**Tables**

| Type | ABREST | BELLERIVE |
| --- | ---: | ---: |
| Household without child | 360 | 1316 |
| Household with one child | 192 | 580 |
| Household with two children | 156 | 444 |
| Household with three children | 48 | 120 |
| Household with four or more children | 16 | 44 |

**Table 1:** Distribution of households according to the number of children for the municipalities Abrest and Bellerive sur Allier. Source: INSEE, French Census data, 1990.

| Type | Proportion |
|---|---|
| Single | 0,2720 |
| Single-parent | 0,0660 |
| Couple | 0,2370 |
| Couple with Children | 0,3640 |
| Complex | 0,0610 |

**Table 2:** Distribution of households according to the type in France. Source: INSEE, 1990.